 \documentclass{pasj01}
\usepackage{natbib}
\usepackage{color}
\usepackage{mediabb}
\Received{$\langle$reception date$\rangle$}
\Accepted{$\langle$acception date$\rangle$}
\Published{$\langle$publication date$\rangle$}

\begin{document}

\title{ALMA Deep Field in SSA22: Survey Design and Source Catalog of a 20\,arcmin$^2$ Survey at 1.1\,mm }
\author{Hideki Umehata,$^{\! 1,2,3}$ Bunyo Hatsukade,$^{\! 3}$ Ian~Smail,$^{\! 4}$ David M.~Alexander,$^{\! 4}$ R.\,J~Ivison,$^{\! 5,6}$ Yuichi~Matsuda,$^{\! 7}$ Yoichi~Tamura,$^{\! 8}$ Kotaro~Kohno,$^{\! 3,9}$ Yuta~Kato,$^{\! 7,10}$ Natsuki~Hayatsu,$^{\! 11,5}$ Mariko~Kubo$^{7}$ and Soh~Ikarashi$^{12}$}
\altaffiltext{1}{RIKEN Cluster for Pioneering Research, 2-1 Hirosawa, Wako-shi, Saitama 351-0198, Japan}
\altaffiltext{2}{The Open University of Japan, 2-11 Wakaba, Mihama-ku, Chiba 261-8586, Japan}
\altaffiltext{3}{Institute of Astronomy, School of Science, The University of Tokyo, 2-21-1 Osawa, Mitaka, Tokyo 181-0015, Japan}
\altaffiltext{4}{Centre for Extragalactic Astronomy, Department of Physics, Durham University, South Road, Durham, DH1 3LE, UK}
\altaffiltext{5}{European Southern Observatory, Karl-Schwarzschild-Str.\ 2, D-85748 Garching, Germany}
\altaffiltext{6}{Institute for Astronomy, University of Edinburgh, Royal Observatory, Blackford Hill, Edinburgh EH9 3HJ, UK}
\altaffiltext{7}{National Astronomical Observatory of Japan, 2-21-1 Osawa, Mitaka, Tokyo 181-8588, Japan}
\altaffiltext{8}{Department of Physics, Nagoya University, Furo-cho, Chikusa-ku, Nagoya 464-8601, Japan}
\altaffiltext{9}{Research Center for the Early Universe, The University of Tokyo, 7-3-1 Hongo, Bunkyo, Tokyo 113-0033}
\altaffiltext{10}{Department of Astronomy, Graduate school of Science, The University of Tokyo, 7-3-1 Hongo, Bunkyo-ku, Tokyo 133-0033, Japan}
\altaffiltext{11}{Department of Physics, Graduate School of Science, The University of Tokyo, 7-3-1 Hongo, Bunkyo, Tokyo, 113-0033, Japan}
\altaffiltext{12}{Kapteyn Astronomical Institute, University of Groningen, P.O.\ Box 800, 9700 AV Groningen,Netherlands}
\email{hideki.umehata@riken.jp}


\maketitle

\begin{abstract}
To search for dust-obscured star-formation activity in the early Universe, it is essential to obtain a deep and wide submillimeter/millimeter map. The advent of the Atacama Large Millimeter/submillimeter Array (ALMA) has enabled us to obtain such maps at sufficiently high spatial resolution to be free from source confusion.
We present a new 1.1\,mm map obtained by ALMA in the SSA22 field.  SSA22 contains a remarkable proto-cluster at $z=3.09$ and is therefore an ideal region to investigate the role of large-scale cosmic web on dust-obscured star formation. The typical 1$\sigma$ depth of our map is 73~$\mu$Jy~beam$^{-1}$ at a 0$^{\prime\prime}$.5 resolution; combined with earlier, archived observations, we map an area of 20 arcmin$^2$ (71 comoving Mpc$^2$ at $z=3.09$).
Within the combined survey area we have detected 35 sources at a signal-to-noise ratio (S/N) $>5$, with flux densities, $S_{\rm 1.1mm}=$0.43--5.6 mJy, equivalent to star-formation rates of $\gtrsim$100--1000 $M_\odot$~yr$^{-1}$ at $z=3.09$, for a Chabrier initial mass function; of these, 17 are new detections.  The cumulative number counts show a factor 3--5$\times$ excess compared to blank fields.  The excess suggests enhanced dust-enshrouded star-formation activity in the proto-cluster on a 10 comoving Mpc scale, indicating accelerated galaxy evolution in this overdense region.
\end{abstract}

\section{Introduction}

One of the major goals of modern astronomy is to uncover obscured star formation across cosmic time and thus to understand the complete cosmic history of star formation.
While star-forming galaxies intrinsically radiate their stellar emission at rest-frame ultraviolet (UV) to optical wavelengths, the bulk of this emission is absorbed and re-radiated in the infrared (IR) since star formation is usually accompanied by a significant amount of dust. In addition to conventional UV to optical observations, we therefore also require observations at far-IR (FIR) to submillimeter/millimeter (submm/mm) wavelengths, which then allows us to trace rest-frame FIR emission from galaxies at all redshifts, essential for a complete understanding of cosmic star-formation history.

A distant galaxy population -- bright at submm wavelengths and showing high dust-obscured star-formation rates (SFRs), the so-called submillimeter galaxies (SMGs) or dusty star-forming galaxies (DSFGs) -- was discovered about twenty years ago (e.g.\ \citealt{1997ApJ...490L...5S}; \citealt{1998Natur.394..241H}; \citealt{1998MNRAS.298..583I}).  More recently, wide-area surveys at (sub)millimeter wavelengths, obtained with ground-based facilities, e.g.\ SCUBA (\citealt{1999MNRAS.303..659H}) and SCUBA-2 (\citealt{2013MNRAS.430.2513H}) on the James Clerk Maxwell Telescope, and AzTEC (\citealt{2008MNRAS.386..807W}) on the Atacama Submillimeter Telescope Experiment (ASTE; \citealt{2004SPIE.5489..763E}; \citeyear{2008SPIE.7012E..08E}), or using SPIRE (\citealt{2010A&A...518L...3G}) on the {\it Herschel Space Observatory}\ (\citealt{2010A&A...518L...1P}),  have discovered an enormous number of such galaxies -- for a recent review, see \citet{2014PhR...541...45C}.  It was discovered that dust-obscured star formation dominates the total SFR budget at earlier times (at least at $z\lesssim3$--4) while it makes a smaller contribution in the present-day Universe (e.g.\ \citealt{2014ARA&A..52..415M}).

Although these wide-area surveys pioneered a new regime of submm/mm astronomy, they suffered from a severe problem: the poor angular resolution of single-dish surveys (typically $15^{\prime\prime}\sim30^{\prime\prime}$) limited sensitivity due to source confusion (e.g.\ \citealt{1998MNRAS.296L..29B}; \citealt{2010A&A...518L...5N}).  In order to obtain an accurate picture of obscured star formation and compare with the views probed by optical and near-IR (NIR) surveys, wide and deep submm/mm mapping -- free from source confusion -- is essential.
The Atacama Large Millimeter/submillimeter Array (ALMA) has delivered a revolution in this regard, providing the capability to map the sky at (sub-)arcsec resolution. There are already several such `ALMA deep surveys' covering contiguous regions of several arcmin$^2$ (e.g.\  \citealt{2016PASJ...68...36H}; \citealt{2016ApJ...833...67W}; \citealt{2017ApJ...835...98U}; \citealt{2017MNRAS.466..861D}; \citealt{2017arXiv171203983M}).
These surveys are beginning to link the obscured/unobscured views of galaxies below the the sensitivity limits of single-dish surveys, though the area and depth of the current ALMA surveys are still very limited.

An important aspect in uncovering dusty star-formation activity is the role of large-scale environment.  A number of works -- based on single-dish surveys -- have claimed a correlation between SMGs and a high density of galaxies in the early Universe (e.g.\ \citealt{2003Natur.425..264S}; \citealt{2009Natur.459...61T}; \citealt{2009ApJ...694.1517D}; \citealt{2011Natur.470..233C}; \citealt{2014MNRAS.440.3462U}; \citealt{2014A&A...570A..55D}), which may correspond to an ancient starburst phase experienced by massive galaxies in proto-clusters -- the massive ellipticals we see in clusters at low redshift (e.g.\ \citealt{1999ApJ...515..518E}; \citealt{2006MNRAS.366..499D}).  It has also been reported that star-forming galaxies in proto-clusters at $z$=2--2.5 tend to be more massive than galaxies in field environments at similar redshifts (e.g.\ \citealt{2005ApJ...626...44S}; \citealt{2013MNRAS.434..423K}), which suggests preferential dust enrichment in proto-clusters together with a trend for massive star-forming galaxies to show high dust-obscured SFRs (e.g.\ \citealt{2017MNRAS.466..861D}).  Thus, unveiling the relationship between dust-obscured star formation and underlying large-scale structures is one of important goals of ALMA deep surveys.

A first attempt to undertake a wide field mm survey with ALMA towards a proto-cluster was published by \citet{2015ApJ...815L...8U} and \citet{2017ApJ...835...98U}. In these papers, we reported a 7 arcmin$^2$ ALMA survey at 1.1\,mm in the $z=3.1$ SSA22 proto-cluster field, which shows highly elevated dust-obscured star-formation activity. In this paper, we present new ALMA observations which significantly expand the contiguous survey area in the SSA22 proto-cluster field to a total of 20 arcmin$^2$, mapping the obscured activity in the proto-cluster across a wide field. This paper is structured as follows: the survey design, observations and data reduction are described in \S\,2; we present the source extraction procedures and final source catalog in \S\,3; we derive source number counts and show the improvement achieved by ALMA compared to a single-dish survey in \S\,4.  Finally, we summarize our paper in \S\,5.  Throughout, we adopt a cosmology with $\Omega_{\rm m}$=0.3, $\Omega_{\rm A}$=0.7, and H$_0$ = 70 km s$^{-1}$ Mpc$^{-1}$. This gives us a physical scale of 7.6~kpc arcsec$^{-1}$ at $z=3.09$.

\section{Observation and Data Reduction}

\subsection{Survey Design}

\begin{figure}
  \begin{center}
\includegraphics[width=8.0cm]{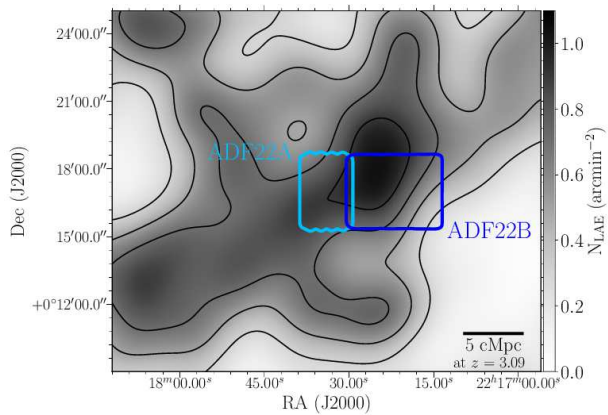}
  \end{center}
  \caption{%
The fields of the ALMA mosaics (``ALMA Deep Field in SSA22'' or ``ADF22'') on the {\it projected} $z=3.09$ LAE surface number density map in SSA22 from \citet{2004AJ....128.2073H}.
The cyan/blue contours show the area we mapped with ALMA in Cycle~2 (``ADF22A'', \citealt{2017ApJ...835...98U}) and Cycle~4 (``ADF22B'', this work) corresponding to the 30\% sensitivity limit of each final mosaic map, respectively.
In total, a 20 arcmin$^2$ contiguous region has been observed by ALMA towards the proto-cluster core.
}%
  \label{fig:lss}
\end{figure}

The remarkable $z=3.09$ proto-cluster in SSA22 was discovered originally as a large over-density of Lyman-break galaxies (LBGs) and Ly$\alpha$ emitters (LAEs) (\citealt{1998ApJ...492..428S}; \citeyear{2000ApJ...532..170S}).  As shown in Fig.~\ref{fig:lss}, \citet{2004AJ....128.2073H} and \citet{2005ApJ...634L.125M} further confirmed a filamentary large-scale structure of LAEs in two/three dimensions -- on a 50 comoving Mpc scale -- on the basis of optical narrow-band imaging and spectroscopy, which makes the field an outstanding and unique laboratory to investigate galaxy formation and evolution within a dense cosmic structure in the early Universe.

As reported in \citet{2017ApJ...835...98U}  (see also \citealt{2015ApJ...815L...8U} and \citealt{2017PASJ...69...45H}), we previously observed a $\approx 2^\prime \times 3^\prime$ region at 1.1\,mm in SSA22 during ALMA Cycle~2.
As shown in Fig.~\ref{fig:lss}, the region was selected because it corresponds to the intersection of the three-dimensional structure traced by LAEs. Through our new ALMA Cycle~4 project (\# 2016.1.00580.S, PI.\ Umehata), we have expanded the 1.1\,mm map (``ALMA Deep Field in SSA22" or ``ADF22'') westward, as shown in Fig.~\ref{fig:lss}, adjacent to the previous survey area. 
We call the two fields observed in Cycles~2 and 4, ADF22A and ADF22B, respectively. Hereafter we use the term ADF22 to represent the total coverage of the two projects\footnote{The survey area, ADF22A, therefore corresponds to the ``ADF22'' area we denoted in our previous papers (\citealt{2015ApJ...815L...8U}; \citeyear{2017ApJ...835...98U}; \citealt{2017PASJ...69...45H})}.

The primary aim of our expanded survey is to cover a much wider area -- about a 3$\times$ increase compared to ADF22A (see \S\,3.3) -- giving us better statistics on the relationship between a larger cosmic structure and the individually detected SMGs in the proto-cluster.  Additionally, the new coverage of ADF22B can trace the level of dust-obscured star-formation activity in different parts of the proto-cluster, as compared with ADF22A.  As shown in Fig.~\ref{fig:lss}, ADF22B encompasses the `projected' densest peak of LAEs, traced by LAEs whose redshift distribution shows two slightly different spikes ($z_{\rm Ly\alpha}\approx3.06$ and $z_{\rm Ly\alpha}\approx3.10$) (Fig.~1 in \citealt{2005ApJ...634L.125M}, see also \citealt{2016ApJ...824L..11T}). By combining the two projects we are therefore able to obtain a 1.1\,mm map not only at the exact intersection of the 3D structure but also covering the filaments of the cosmic large-scale structure (or `cosmic web'). The 1~mm map is expected to be also an essential foundation for a complete spectroscopic survey of SMGs, using ALMA to detect their spectral lines.

SSA22 is also covered by a rich array of multi-wavelength imaging datasets of this field, including {\it Chandra} ACIS (\citealt{2009ApJ...691..687L}; \citeyear{2009MNRAS.400..299L}), Subaru/Suprime-Cam (\citealt{2004AJ....128.2073H}), Subaru/MOIRCS (\citealt{2012ApJ...750..116U}), {\it Spitzer}/IRAC and MIPS (\citealt{2009ApJ...692.1561W}), {\it Herschel}/SPIRE (\citealt{2016MNRAS.460.3861K}), the 870~$\mu$m ALMA pointings (\citealt{2016MNRAS.461.2944A}), JVLA (\citealt{2004ApJ...611..732C}; \citealt{2017ApJ...850..178A}), and optical/NIR spectroscopic catalogs (e.g.\  \citealt{1998ApJ...492..428S}; \citealt{2013ApJ...765...47N}; \citealt{2014ApJ...795...33E}; \citealt{2015MNRAS.450.2615S}; \citealt{2015ApJ...799...38K}; \citeyear{2016MNRAS.455.3333K}). The combination of these rich ancillary datasets make the ADF22 region exceedingly rare and valuable. 

\subsection{Observations}

\begin{figure*}
  \begin{center}
\includegraphics[width=16.0cm]{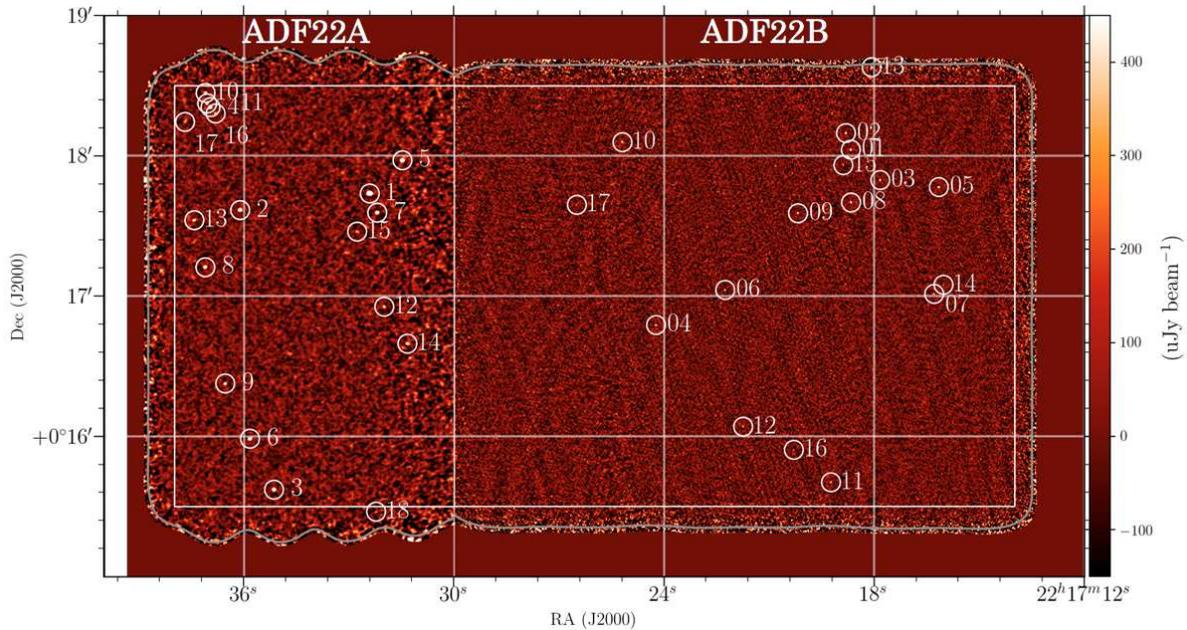}
  \end{center}
  \caption{%
ALMA 1.1\,mm map of ADF22. The western side shows the ADF22B map and the inner white rectangle shows a $4^{\prime}\times3^{\prime}$ region. Gray contours show the map area where the primary beam attenuation is less than 30\%, where source extraction was performed. The white circles show the detected ALMA sources with labels for each source ID (details are described in \S\,3). For reference, the 1$^{\prime\prime}$-tapered ADF22A map and source positions (\citealt{2017ApJ...835...98U}) are also showcased on the eastern side in a similar manner (the inner rectangle shows a $2^{\prime}\times3^{\prime}$ region). In total, a contiguous 20~arcmin$^2$  region is covered by ALMA. Note that although the two maps are combined here for convenience of display, all analyses were done separately using each map.
}%
  \label{fig:almamap}
\end{figure*}

We split our target field, an around $4^\prime \times 3^\prime$ region (Fig.~\ref{fig:lss}), into three sub-regions (Tile a, Tile b, and Tile c) due to a limit on the maximum number of pointing for ALMA mosaic observations allowed by the observatory (i.e.\ 150 pointings at most). The central coordinates of the three tiles are ($\alpha$,$\delta$) = (22$^{\rm h}$17$^{\rm m}$27.4$^{\rm s}$, $+00^\circ17^\prime00^{\prime\prime}$), (22$^{\rm h}$17$^{\rm m}$22.0$^{\rm s}$, $+00^\circ17^\prime00^{\prime\prime}$), (22$^{\rm h}$17$^{\rm m}$16.7$^{\rm s}$, $+00^\circ17^\prime00^{\prime\prime}$) for Tile a, Tile b, and Tile c, respectively.
Each tile was designed to have a 133-point mosaic with Nyquist sampled spacings.
Consequently the observations of ADF22B were divided into nine execution blocks (EBs).
Each observation was executed on 2017 July 8--12 with 42--43 available 12m antennas in the C40-5 configuration.
All EBs were carried out with a same array configuration and therefore we achieved a uniform synthesized beam size across the whole ADF22B field, though the maximum baseline, 2.6~km, was much longer than we originally requested. The weather conditions were excellent or good (precipitable water vapor (PWV) of 0.3--1.0~mm). The exposure time per pointing was 1 min so that the total amount of on-source time was 399 min. The total observation time including overheads was 11.3 hours.

The central observing frequency of 263~GHz (1.14~mm) was selected to be very close to that of our previous AzTEC/ASTE survey, 270 GHz, allowing us to compare the flux densities of sources between the two maps (we will discuss this in \S\,4.3).  We used the TDM correlator mode.  The central frequencies of the four spectral windows were 254.0, 256.0, 270.0 and 272.0 GHz, respectively. The correlator was set up to target each spectral window of 1.875 GHz bandwidth at 15.6~MHz ($\sim$ 20 km s$^{-1}$) channel spacing in each sideband, each with 128 dual-polarization channels. Another advantage of this frequency set-up is the coverage it provides of the redshifted $^{12}$CO(9--8) line ($\nu_{\rm rest}=$1036.912 GHz) at $z\sim3.09$, which was demonstrated by the detection of one proto-cluster SMG in ADF22A (\citealt{2017PASJ...69...45H}). We note that the frequency set-ups are exactly the same as those in ADF22A, which allows us to combine the two datasets relatively easily.

The quasar, J2226+0052,  was observed regularly for amplitude and phase calibration.
The absolute flux density scale was set using the quasar, J2226+0052.
The bandpass calibration was performed using J2148+0657.
The absolute flux density accuracy is evaluated to be within 10\%.
This uncertainty in the absolute flux density calibration is not included in the following analyses and discussions.

\subsection{Data Reduction}

Data reduction was performed using the Common Astronomy Software Application ({\sc casa}).
The calibration process was carried out using the {\sc casa} version of the ALMA pipeline (Pipeline-Cycle4-R2-B) for each tile.
A continuum map of ADF22B was created from the calibrated data using the task {\sc tclean} with the {\sc casa} version 5.1.1.
The $u v$ data for each pointing for all the three tiles was first combined into a single $u v$ dataset.
To enhance the speed of the imaging process, the data were averaged in time with a sampling of 10.296~s.
We then Fourier-transformed the combined data to create a single ``dirty'' map with natural weighting to achieve the maximum possible sensitivity, by setting the gridder mode to ``mosaic'' and the deconvolver to ``hogbom''.
Without any taper, the resultant size of the synthesized beam is $0.26^{\prime\prime}\times0.20^{\prime\prime}$ (P.A.\ $-59$ deg).
As we mentioned earlier, this angular resolution was higher than we had requested ($\approx0.5^{\prime\prime}-1.5^{\prime\prime}$), hence we applied a taper to achieve an angular resolution closer to our request, so as not to resolve any sources significantly, though such a procedure forces us to lose some point-source sensitivity.
As a compromise, we adopted the taper parameter, ``uvtaper = 0.5 arcsec''. 
After we measured the r.m.s.\ noise level across the whole dirty map, we repeated the clean process down to 3$\sigma$, putting a tight clean box around each 5$\sigma$ source.
The resulting synthesized beam size is $0.53^{\prime\prime}\times0.52^{\prime\prime}$ (P.A. $-54$ deg).
The typical 1$\sigma$ depth measured over a large central area of the map is 73 $\mu$Jy beam$^{-1}$.
The final maps are shown in Fig.~\ref{fig:almamap}.
The survey area where attenuation of the primary beam is less than 30\% is 13 arcmin$^2$.

Considering the fact that our ADF22 survey is composed of two fields, ADF22A and ADF22B, one concern here is the homogeneity of the two maps.  As we showed in \citet{2017ApJ...835...98U}, ADF22A has typical noise levels of 75 $\mu$Jy beam$^{-1}$ and 60 $\mu$Jy beam$^{-1}$ at resolutions of 1 $^{\prime\prime}$ and 0.7 $^{\prime\prime}$, respectively, though the latter map covers only about 80\% of ADF22A. Therefore in general the two fields have comparable qualities, and almost uniform sensitivity is provided across a 20 arcmin$^2$ sky area toward the $z=3.09$ proto-cluster. This is the largest contiguous ALMA map in a known proto-cluster field so far.  We note that our intrinsic angular resolutions are not exactly uniform across ADF22 and hence sensitivity to relatively extended sources may not be homogeneous. 

\section{Source Catalog}

\subsection{Source Extraction}

\begin{figure}
  \begin{center}
\includegraphics[width=8.0cm]{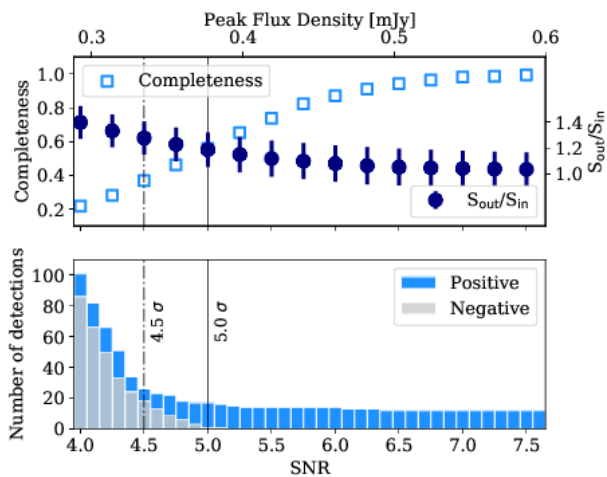}
  \end{center}
  \caption{%
  (top)
Completeness and the input/output flux density ratio as a function of  signal-to-noise ratio (SNR), both of which are measured through a series of simulations to inject and extract mock sources (for detail, see \S\,3.2). The upper axis shows the corresponding peak flux densities, assuming a typical r.m.s.\ level, $1\sigma = 73 \mu$Jy.
Completeness goes down as SNR decreases at SNR$<6$, and reaches $\sim$55\% at SNR=5. The cumulative completeness above SNR is $\sim$92\%, which suggests that one or two real sources may be missed.
The input/output flux density ratio indicates that the measured flux density of $5\sigma$ sources might be over-estimated (by $\approx$20\%, on average) due to flux boosting.
(bottom)  
Cumulative number of positive and negative peaks as a function of SNR for the ADF22B map.
The solid vertical line shows our adopted detection limit ($5\sigma$).
There is just one negative peak above 5$\sigma$, indicating a false positive rate of $\lesssim5$\%.
The dash-dot line shows the threshold for a source candidate which is classified into a supplementary catalog ($4.5\sigma$). 
}%
  \label{fig:mq}
\end{figure}

Source extraction procedures for the ADF22B map were performed using the source-finding algorithm, {\sc aegean} v2.0b191 (\citealt{2012MNRAS.422.1812H}) as was the case with our previous survey (\citealt{2017ApJ...835...98U}). First a sensitivity map was constructed using the {\sc aegean} task {\sc bane}, calculating the standard deviation for a pixel
 using the surrounding $100\times100$ pixels (since each pixel is $0.1^{\prime\prime}\times0.1^{\prime\prime}$ in size, a $10.0^{\prime\prime}\times10.0^{\prime\prime}$ region was considered for each) on an image prior to the correction of primary beam response. During the calculation, we applied 3$\sigma$ clipping, so as not to be affected by bright sources. The background fluctuation on larger scales was also estimated by {\sc bane}.

The sensitivity and background maps allowed us to calculate a signal-to-noise (S/N) map which is used for source detection. Using {\sc aegean}, we extracted both positive and negative peaks above S/N=4. The cumulative number counts of positive and negative peaks as a function of signal-to-noise ratio (SNR) is shown in the bottom panel of Fig.~\ref{fig:mq}. Consequently 17 sources are found above S/N=5. Since the number of negative peaks above the equivalent threshold is just one, we adopt these 17 sources as secure detections.
Although there are further candidates at SNR$<5$, the increasing number of negative detections gives us a warning that these candidates with moderate SNR may be heavily contaminated by false detections.
We list nine sources with SNR = 4.5--5 in a supplementary catalog. There is an equivalent or larger number of negative detections in this SNR range and therefore we do not use them in following discussion. A much deeper mm map is required to extract fainter sources securely.

\subsection{Completeness and flux boosting}

A suite of simulations was executed to evaluate the completeness and flux boosting effect on the flux density measurements, following the recipe described in \citet{2016PASJ...68...36H} and B.~Hatsukade et al. (2018, in preparation).
In the simulation, mock sources are created, assumed a symmetric two-dimensional Gaussian profile with a range of FWHM ($0.1^{\prime\prime}$--$0.5^{\prime\prime}$, in steps of 0$^{\prime\prime}$.1) and SNR (3.0--10.0, in steps of 0.25).
The mock sources are injected into the image map prior to primary beam correction, 1000 times for each SNR bin. The input position is randomly selected, avoiding any $5\sigma$ sources, and source extraction was again performed using {\sc aegean}, as described in \S\,3.1.

The completeness was calculated as the recovery rate of the injected mock sources. A source is recognized to be recovered if it is detected with a certain level of significance ($>4\sigma$)  at a position close to the originally injected mock source ($<1.0^{\prime\prime}$). The upper panel of Fig.~\ref{fig:mq} shows the completeness as a function of the intrinsic SNR of mock sources. Completeness is below 90\% at SNR$\lesssim$6 and reduces as SNR decreases. At SNR=5, which is the detection threshold, the completeness is estimated to be $\sim$55\%. The integrated completeness for the 17 sources is $\approx$92\%, considering the SNR distribution (Table~\ref{table:catalog}). This indicates that  1--2 relatively faint sources would be missed due to random noise fluctuations at our adopted SNR threshold.
We correct for this effect when calculating source number counts in \S\,4.1. 

It is known that the flux densities of sources in a signal-to-noise-limited catalog tend to be boosted due to random noise fluctuations and the shape of the source count distribution (e.g.\  \citealt{1998PASP..110..727H}). To test the flux boosting effect,  we examined the ratio of two flux densities for each source; the value which is intrinsically given in creating a mock source and the measurement on the image. Sources with SNR$\sim$5 may suffer from $\approx$20\% artificial increase of flux density due to this effect (the top panel of Fig.~\ref{fig:mq}).
We correct this effect in calculating source number counts (\S\,4.1).
This effect is not considered in cataloging (\S\,3.3) since one-to-one correspondence is not clear and dispersion is large.
But this result provides a caveat on the flux density measured with relatively low SNR. 

\subsection{Measurement of Source Properties}

\renewcommand{\arraystretch}{0.6} 
\begin{table*}
  \tbl{ADF22 Source Catalog}{%
  \begin{tabular}{cccccccc}
  \hline\noalign{\vskip3pt}
  (1) & (2) & (3) & (4) & (5) & (6) & (7) & (8) \\
  ID & R.A.  & Dec.  & S/N & $S_{\rm 1.14mm}$ & Deconvolved beam & P.A. & $z_{\rm spec}$ \\
      &  (J2000)         &    (J2000)      &        &  [mJy]               &   [arcsec$\times$arcsec] & [deg] & \\
  \hline\noalign{\vskip3pt}
   \multicolumn{8}{c}{ADF22B Main source catalog}   \\
    \hline\noalign{\vskip3pt}
ADF22B.1  &  22:17:18.66  &   +00:18:02.6  &   25.5  &   2.18 $\pm$ 0.13   &  (0.12  $\pm$ 0.02) $\times$ (0.07 $\pm$ 0.06)  & 29 $\pm$ 45 & 3.067\\  
ADF22B.2  &  22:17:18.80  &   +00:18:097  &   24.8  &   2.20 $\pm$ 0.09   &   (0.31 $\pm$ 0.03)  $\times$ (0.17 $\pm$ 0.03) & 178 $\pm$ 9 &...\\ 
ADF22B.3  &  22:17:17.82  &   +00:17:49.6  &   16.2  &   1.38 $\pm$ 0.09   &   (0.24  $\pm$ 0.04) $\times$ (0.12  $\pm$ 0.08)  & 12 $\pm$ 19 &...\\  
ADF22B.4  &  22:17:24.23  &   +00:16:47.6  &   15.5  &   1.45 $\pm$ 0.11   &   (0.19  $\pm$ 0.03) $\times$ (0.15  $\pm$ 0.03)  & 119 $\pm$ 67 &...\\  
ADF22B.5$^\S$  &  22:17:16.15  &   +00:17:46.5  &   14.5  &   1.61 $\pm$ 0.12   &   (0.29  $\pm$ 0.06) $\times$ (0.25  $\pm$ 0.06)   & 138 $\pm$ 75 &...\\   
ADF22B.6  &  22:17:22.25  &   +00:17:02.6  &   14.3  &   1.13 $\pm$ 0.05   &   (0.15  $\pm$ 0.05) $\times$ (0.14  $\pm$ 0.06) & 105 $\pm$ 79 &...\\   
ADF22B.7  &  22:17:16.29  &   +00:17:00.8  &   13.8  &   1.04 $\pm$ 0.10   &  (0.14  $\pm$ 0.04) $\times$ (0.03  $\pm$ 0.06)   & 152 $\pm$ 33 &...\\   
ADF22B.8  &  22:17:18.66  &   +00:17:40.1  &   13.5  &   1.56 $\pm$ 0.11   &  (0.48  $\pm$ 0.07) $\times$ (0.25  $\pm$ 0.06) & 12 $\pm$ 10 &...\\ \ 
ADF22B.9$^\dagger$  &  22:17:20.17  &   +00:17:35.4  &   13.5  &   1.33 $\pm$ 0.11   &  (0.34  $\pm$ 0.06) $\times$ (0.19   $\pm$ 0.07) & 174 $\pm$ 21 &...\\ 
ADF22B.10  &  22:17:25.18  &   +00:18:05.8  &   10.8  &   1.49 $\pm$ 0.13   &  (0.31  $\pm$  0.06) $\times$ (0.16  $\pm$ 0.08) & 43 $\pm$ 19 & 2.319\\   
ADF22B.11  &  22:17:19.23  &   +00:15:40.3 &   9.2  &   0.86 $\pm$ 0.08      &  --- &  --- & 3.101\\
ADF22B.12  &  22:17:21.74  &   +00:16:04.2  &   9.0  &   0.81 $\pm$ 0.16   &  --- &  --- &...\\ 
ADF22B.13  &  22:17:18.07  &   +00:18:37.9  &   6.4  &   2.30 $\pm$ 0.31   &  --- &  --- &...\\ 
ADF22B.14  &  22:17:16.01  &   +00:17:04.8  &   6.0  &   0.43 $\pm$ 0.08   &  --- &  --- &...\\ 
ADF22B.15  &  22:17:18.88  &   +00:17:56.1  &   5.3  &   0.73 $\pm$ 0.18   &  --- &  --- &...\\ 
ADF22B.16  &  22:17:20.29  &   +00:15:54.1  &   5.1  &   0.44 $\pm$ 0.12   &  --- &  ---&...\\ 
ADF22B.17  &  22:17:26.48  &   +00:17:39.0  &   5.1  &   0.67 $\pm$ 0.13   &  --- &  --- & 2.168 \\
  \hline\noalign{\vskip3pt}
  \multicolumn{8}{c}{ADF22A Main Source Catalog}   \\
    \hline\noalign{\vskip3pt}
ADF22A.1$^\ddag$ & 22:17:32.41 & +00:17:43.8  &  58.1   &   5.60 $\pm$ 0.13 & (0.85 $\pm$ 0.04)  $\times$ (0.33 $\pm$ 0.03)  & 82 $\pm$ 2 & 3.092\\ 
ADF22A.2  & 22:17:36.11 & +00:17:36.7 & 31.8   &   2.02 $\pm$ 0.02  & (0.29 $\pm$ 0.05) $\times$  (0.05 $\pm$ 0.11)  & 6 $\pm$ 21 & ...\\  
ADF22A.3  & 22:17:35.15 & +00:15:37.2 &  27.0   &   1.89 $\pm$ 0.04 & (0.45 $\pm$ 0.05) $\times$  (0.18 $\pm$ 0.08) & 18 $\pm$ 20  & 3.096\\ 
ADF22A.4$^\ddag$  & 22:17:36.96 & +00:18:20.7 & 26.6   &   1.95 $\pm$ 0.05 & (0.25 $\pm$ 0.05) $\times$ (0.04 $\pm$ 0.11)  & 78 $\pm$ 9 & 3.091\\
ADF22A.5  &  22:17:31.48 & +00:17:58.0 & 20.3   &   2.43 $\pm$ 0.20 & (0.57 $\pm$ 0.06)  $\times$ (0.26 $\pm$ 0.06)  & 29 $\pm$ 86 & ...\\ 
ADF22A.6$^{\dagger}$$^{\ddag}$$^\S$  &  22:17:35.83 & +00:15:59.0  & 19.1   &   1.45 $\pm$ 0.09 & (0.38 $\pm$ 0.08) $\times$ (0.30 $\pm$ 0.14) & 21 $\pm$ 64 & 3.089 \\
ADF22A.7$^\ddag$  &  22:17:32.20 & +00:17:35.6  & 18.7   &   1.65 $\pm$ 0.07 & (0.32 $\pm$ 0.08) $\times$ (0.22 $\pm$ 0.10)  & 7 $\pm$ 53 & 3.097\\
ADF22A.8  &  22:17:37.11 & +00:17:12.3  & 15.0   &   1.19 $\pm$ 0.06 & (0.26 $\pm$ 0.07) $\times$  (0.18 $\pm$ 0.05)  & 84 $\pm$ 8 & 3.090\\
ADF22A.9$^\ddag$$^\S$  & 22:17:36.54 & +00:16:22.6  &  12.8   &   0.82 $\pm$ 0.08 & (0.28 $\pm$ 0.14) $\times$ (0.26 $\pm$ 0.17)  & 106 $\pm$ 89 & 3.095\\
ADF22A.10  & 22:17:37.10 & +00:18:26.8  &  9.8   &   0.72 $\pm$ 0.04 &  --- &  --- & ...\\
ADF22A.11  &  22:17:37.05 & +00:18:22.3  &  9.5   &   0.79 $\pm$ 0.05 &  --- &  --- & 3.093\\
ADF22A.12$^\dagger$$^\ddag$$^\S$  & 22:17:32.00 & +00:16:55.4  & 8.8   &   0.63 $\pm$ 0.03   &  --- &  ---& 3.091 \\
ADF22A.13  & 22:17:37.42 & +00:17:32.4  & 8.0   &   0.79 $\pm$ 0.05   &  --- &  --- & ...\\
ADF22A.14  & 22:17:31.34 & +00:16:39.6  & 7.5   &   0.98 $\pm$ 0.13    &  --- &  --- & ...\\
ADF22A.15  & 22:17:32.77 & +00:17:27.5  & 5.3   &   0.50 $\pm$ 0.08  &  --- &  --- & ...\\
ADF22A.16  & 22:17:36.81 & +00:18:18.0  & 5.3   &   0.56 $\pm$ 0.07 &  --- &  --- & 3.085\\
ADF22A.17  & 22:17:37.69 & +00:18:14.4  &   5.1   &   0.60 $\pm$ 0.09  &  --- &  --- & ...\\
ADF22A.18$^\ddag$$^{\triangleright}$ & 22:17:32.23 & +00:15:27.8  &  5.3  &   0.44 $\pm$ 0.05 &  --- &  --- & 2.105\\
\hline\noalign{\vskip3pt}
  \end{tabular}}\label{table:catalog}
  \begin{tabnote}
 Both ADF22B sources (this work) and ADF22A sources (\citealt{2017ApJ...835...98U}) are listed.
(1) ID in this paper. 
(2) Right Ascension in the wcs system.
(3) Declination in the wcs system.
(4) Signal-to-noise (S/N).
(5) Integral flux density measured with {\sc casa}/{\sc imfit}.
(6) Deconvolved source size and position angle via two-dimensional Gaussian fitting by {\sc casa}/{\sc imfit} on the non-tapered map. Only sources with SNR$>$10 are listed, since relatively low SNR hampers us measuring the size correctly. We note that the natural beam size for ADF22B is $\sim3\times$ better than ADF22A and so the measurements for formers may be relatively reliable (\S\, 3.3).
(7) Spectroscopic redshift (${z_{\rm spec}}$). The redshift information for ADF22B sources are from our recent $K$-band spectroscopy taken with Keck/MOSFIRE (H. Umehata et al. in preparation). That of ADF22A sources are summarized in \citealt{2017ApJ...835...98U}.
$^\dagger$: SMGs which appears to be associated with LABs (\citealt{2004AJ....128..569M}).
$^\ddag$: SMGs with X-ray AGNs (\citealt{2009MNRAS.400..299L})
$^\S$: SMGs with 870 $\mu$m detection (\citealt{2016MNRAS.461.2944A})
$^{\triangleright}$: ADF22A sources selected from a $0.7^{\prime\prime}$ map (`DEEP/HIRES' map; see \citealt{2017ApJ...835...98U}). The other ADF22A sources are selected from a  $1^{\prime\prime}$ map (`WIDE/LORES' map).
 \end{tabnote}
\end{table*}
\renewcommand{\arraystretch}{1}

For sources selected by {\sc aegean}, we performed two-dimensional Gaussian fitting using the {\sc imfit} task of {\sc casa} on the map that was applied for the correction of the primary beam response (Fig.~\ref{fig:almamap}). The measured properties of ADF22B sources are summarized in Table~\ref{table:catalog}, together with those of ADF22A sources (\citealt{2017ApJ...835...98U}).
After combining the two fields, ADF22A and ADF22B, the main source catalog consists of 19 sources with SNR$>$10 and 16 sources with SNR=5--10. 
The projected, surface number density of a 5$\sigma$ ALMA source is $\rho \sim1.8$ arcmin$^{-2}$, which surpasses those of other exiting ALMA deep surveys (e.g., $\rho \sim1.1$ arcmin$^{-2}$ in \citealt{2017MNRAS.466..861D}, see also \citealt{2016PASJ...68...36H}; \citealt{2016ApJ...833...68A}; \citealt{2017arXiv171203983M})
, though it depends on various parameters such as sensitivity, angular resolution, and observed wavelengths.
In \S\,4.1, we will make a comparison between our proto-cluster survey and blank-field searches, using source number counts. An additional 21 sources with SNR$\sim$4--5 comprise a supplementary source catalog in Appendix (Table~3), though they are not considered in what follows.
We also list available $z_{\rm spec}$ in Table~\ref{table:catalog} (\citealt{2017ApJ...835...98U} and references therein; H.~Umehata et al., in preparation). We also note that all of the SMGs do not overlap with known LAEs/LBGs at $z=3.09$ (\citealt{2003ApJ...592..728S}; \citealt{2012AJ....143...79Y}) except for one QSO (ADF22A.9) and another X-ray AGN host (ADF22A.6).

The 35 ALMA sources have flux densities, $S_{\rm 1.1mm}=$0.43--5.60 mJy, 19 of which have $S_{\rm 1.1mm}>1$~mJy. Around 40\% of them represent the `sub-mJy' population in the 1.1\,mm regime.
The flux densities provide rough estimates of IR luminosity using a template set of spectral energy distribution (SED) for ALMA-identified SMGs (\citealt{2017ApJ...840...78D}). The range of the flux densities corresponds to the median IR luminosity, $L_{\rm IR[8-1000\mu m]}=0.8_{-0.3}^{+0.2}\times 10^{12}$ -- $1.0_{-0.4}^{+0.3}\times 10^{13}L_\odot$ (or SFR$_{\rm IR}=80_{-40}^{+30}$ -- $1040_{-460}^{+340}$~$M_\odot$~yr$^{-1}$ for a Chabrier IMF; \citealt{2003PASP..115..763C}) at $z=3.09$.
This primitive estimate suggests that the ALMA sources found in ADF22 are dominated by `ULIRGs', while our ALMA observations are beginning to uncover `LIRGs' at high redshift to a certain degree. Hereafter we call these individual ALMA sources SMGs in this paper.

Source sizes are also measured through two-dimensional Gaussian fitting with {\sc casa/imfit}. The measured parameters of deconvolved beam size, the major/minor axis and position angle, are also stored for SMGs with SNR$>10$ in Table 1. For ADF22B SMGs, we utilize the non-tapered map to make use of the original, higher angular resolution.
Consequently all of the bright ADF22B SMGs are resolved on the map. The median deconvolved major axis is $0.27_{-0.12}^{+0.05}$ arcsec (equivalent to 2.1$_{-0.9}^{+0.4}$~physical kpc at $z=3.09$), where the errors were calculated as a a 95\% bootstrap confidence interval.
This estimate is consistent with that of ADF22A SMGs ($0.32_{-0.06}^{+0.13}$ arcsec; \citealt{2017ApJ...835...98U}), which were measured on the ADF22A map (the synthesized beam is $0.70^{\prime\prime}\times0.59^{\prime\prime}$).
These measurements for ADF22 SMGs are on average comparable to previous work for ALMA-identified SMGs originally discovered by single-dish telescopes (e.g.\  $0.30\pm0.04$ arcsec; \citealt{2015ApJ...799...81S}, see also \citealt{2015ApJ...810..133I}; \citealt{2016ApJ...833..103H}).
Considering the fact that the majority of ADF22A sources were at $z_{\rm spec}\approx3.09$ (\citealt{2017ApJ...835...98U}), as in the case of two ADF22B SMGs, the source size measurement implies no strong environmental dependence on the size of a dusty starburst core at the current resolution and sensitivity, as claimed in \citet{2017ApJ...835...98U}.

\section{Discussion}

\subsection{Number Counts}

Source number counts have been widely used to describe the nature of the galaxy population discovered by submm/mm surveys (e.g.\  \citealt{1999MNRAS.302..632B}; \citealt{2006MNRAS.372.1621C}; \citealt{2012MNRAS.423..575S}; \citealt{2013ApJ...769L..27H}; \citealt{2017MNRAS.465.1789G}).
Recent work using ALMA has improved the picture thanks to the advantages of its sensitivity and angular resolution.
As predicted by identification work using radio interferometry (e.g.\ \citealt{2007MNRAS.380..199I}), ALMA follow-up studies of bright single-dish-selected submm/mm sources suggest that multiple SMGs are often encompassed within a relatively large beam of the single-dish surveys and the number counts need to be corrected for this effect (e.g.\  \citealt{2013MNRAS.432....2K}; \citealt{2013ApJ...768...91H}; \citealt{2015ApJ...807..128S})\footnote{In this paper, we use the term ``SMGs'' only for individual submm/mm galaxies identified by an interferometer such as ALMA. Sources selected by single-dish telescopes are instead termed ``submm sources''.}. 
Deep surveys with ALMA
have uncovered the fainter regime ($S_{\rm 1.1mm}\lesssim1$~mJy, e.g.\  \citealt{2016ApJ...833...68A}; \citealt{2017MNRAS.466..861D}) and some work argues that such faint submm/mm galaxies detected by ALMA individually can account for the majority or all of far-IR extragalactic background light (EBL, e.g.\ \citealt{2016PASJ...68...36H}).

The galaxy density environment, which represents the cosmic large-scale structure, is one important parameter influencing the number counts of submm/mm sources. \citet{2011MNRAS.415.3831A} suggested that foreground galaxies and/or galaxy clusters might elevate the (single-dish) source counts in the bright regime ($S_{\rm 1.1mm}\gtrsim$5 mJy). \citet{2017MNRAS.465.1789G} also reported a marginal ($2\sigma$) excess of SCUBA2 source counts in GOODS-N, which encompasses a known proto-cluster at $z=4.1$. Thus number counts have the potential to indicate large-scale structures related to submm/mm sources without redshift information. This viewpoint is also of importance in ALMA deep surveys, which may be sensitive to remarkable (proto-)clusters or, more generally, cosmic variance.  In our previous paper (\citealt{2017ApJ...835...98U}), we reported the excess of number counts in ADF22A, which is undoubtedly caused by genuine proto-cluster members (\citealt{2015ApJ...815L...8U}). The new survey area, ADF22B, allows us to examine the source number counts in a 3$\times$ larger area toward the $z=3.1$ proto-cluster.

We calculated the cumulative number counts and errors with a recipe presented in \citet{2016PASJ...68...36H} and B.~Hatsukade~et~al., in preparation. As described in \S\,3.3, there are seventeen sources above 5$\sigma$ and all of them are considered. We exclude other less securely detected source candidates since vague thresholds can warp the intrinsic results and lead us to over-estimate the counts (as outlined by \citealt{2016ApJ...822...36O} and \citealt{2017ApJ...835...98U}). Since there is only one negative `source' in the ADF22B map, the effect of false detections is not considered in this paper.  The effect of the completeness and flux boosting is corrected on the basis of our simulation, shown in Fig.~\ref{fig:mq}. Uncertainties in each bin are estimated by bootstrapping. 


\begin{figure}
  \begin{center}
\includegraphics[width=8.0cm]{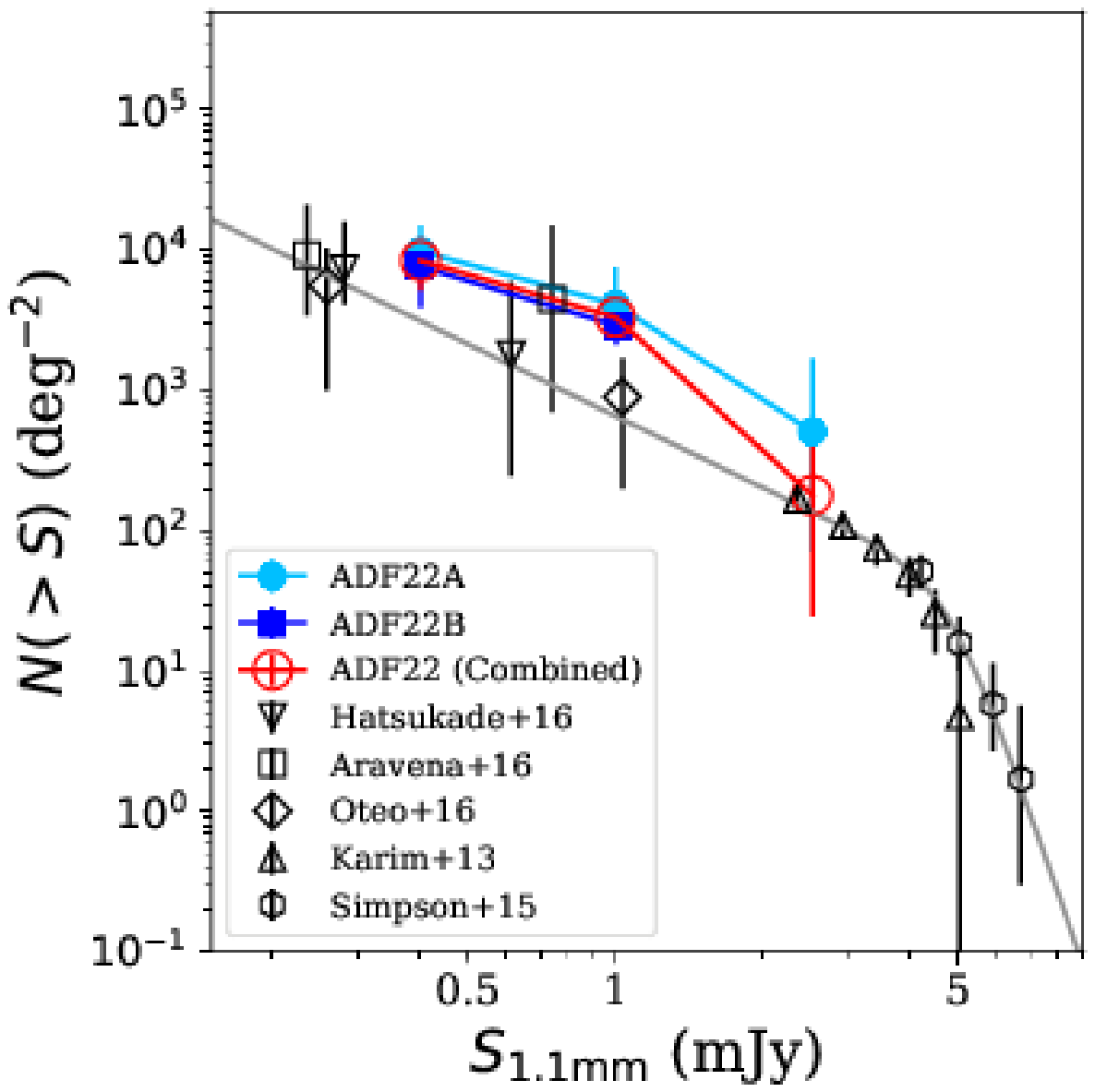}
  \end{center}
  \caption{%
     Cumulative number counts in ADF22 and other ALMA fields. Cyan and blue filled symbols represent the counts in ADF22A and ADF22B, respectively. Combined ADF22 counts are shown with red open circles.
     For comparison, counts in blank fields (\citealt{2016PASJ...68...36H}, \citealt{2016ApJ...833...68A}, \citealt{2016ApJ...822...36O}, \citealt{2015ApJ...807..128S}, and \citealt{2013MNRAS.432....2K}) are also displayed with black symbols. The flux densities of the counts are scaled to 1.1\,mm flux densities assuming a modified black body ($S_{\rm 1.1\,mm}$/$S_{\rm 870\mu m} = 0.56$, $S_{\rm 1.1\,mm}$/$S_{\rm 1.2~mm}=1.29$).
    The grey curve is the best-fit function of a double-power law for the counts in blank fields.
    The counts in ADF22 show a factor of three to five excess compared to the best-fit function, which is expected to be caused by $z=3.09$ proto-cluster members.
      }%
  \label{fig:counts}
\end{figure}

\begin{table}
  \tbl{Cumulative number counts in ADF22.
    }{%
  \begin{tabular}{ccc}
  \hline\noalign{\vskip3pt}
  Flux Density, $S_{\rm 1.1mm}$ & Source, N($>$S) & Number Counts, N($>$S) \\
  (mJy) & & (  $\times10^3$ deg$^{-2}$) \\
  \hline\noalign{\vskip3pt}
  \multicolumn{3}{c}{ADF22 (Combined)}   \\
    \hline\noalign{\vskip3pt}
0.40--1.00 & 32 & $8.4^{+4.2}_{-3.2}$	 \\
1.00--2.52 & 20 & $3.3^{+1.7}_{-1.0}$	 \\
2.52--6.34 & 1 & $0.18^{+0.42}_{-0.15}$ 	 \\
  \hline\noalign{\vskip3pt}
  \multicolumn{3}{c}{ADF22B}   \\
    \hline\noalign{\vskip3pt}
0.40--1.00 & 15 & $7.7^{+3.7}_{-3.7}$ 	 \\
1.00--2.52 & 11 & $2.9^{+0.8}_{-0.8}$	 \\
2.52--6.34 & 0 & ---	 \\
    \hline\noalign{\vskip3pt}
  \multicolumn{3}{c}{ADF22A}   \\
    \hline\noalign{\vskip3pt}
0.40--1.00	& 17 & $9.8^{+5.0}_{-2.2}$ 	 \\
1.00--2.52 & 8 & $4.1^{+3.4}_{-1.4}$	 \\
2.52--6.34 & 1 & $0.5^{+1.2}_{-0.4}$ 	 \\
\hline\noalign{\vskip3pt}
  \end{tabular}}\label{tb:count}
 \begin{tabnote}
Source counts in ADF22A and ADF22B are combined, weighted by the survey area.
The 1$\sigma$ uncertainties are from the bootstrap test for the ADF22B counts.
ADF22A source counts are from \citet{2017ApJ...835...98U}.
The flux bin minima are plotted in Fig.~\ref{fig:counts} for all cases.
 \end{tabnote}
\end{table}

The results are shown in Fig.~\ref{fig:counts}. As summarized in Table~\ref{tb:count}, the ADF22B counts are calculated with the same bins used for the ADF22A counts, as presented in \citet{2017ApJ...835...98U}. We combined the counts in ADF22A and ADF22B, weighting by the survey area, so as to describe the overall counts across the 20~arcmin$^2$ ADF22 area.

While the source counts in typical environments as a gauge to investigate the role of the large-scale structure on the source counts, there are still uncertainties, especially in the regime of $S_{\rm 1.1mm}\lesssim1$~mJy where the counts can show up to a $\sim$1 dex deviation (e.g.\  \citealt{2013ApJ...769L..27H}; \citealt{2016ApJ...833...68A}; \citealt{2016PASJ...68...36H}; \citealt{2016ApJ...822...36O}, \citealt{2016ApJS..222....1F}, \citealt{2017arXiv171203983M}). In order to derive relatively reliable source counts in blank fields, we focus the counts from \citet{2016ApJ...833...68A}, \citet{2016ApJ...822...36O}, and \citet{2016PASJ...68...36H} at $S_{\rm 1.1mm}\lesssim1$~mJy, since they are relatively free from cosmic variance or other biases (the first one utilizes calibrator fields and the latter two use contiguous mapping), and use only 5$\sigma$ sources in a flux regime which should be not contaminated by false detections\footnote{While \citet{2016PASJ...68...36H} reported counts on the basis of a $4\sigma$ threshold in their table, here we adopted the counts calculated only from 5$\sigma$ sources, which were shown in Fig.~6 in the paper.}. Combined with bright source counts from \citet{2015ApJ...807..128S} and \citet{2013MNRAS.432....2K}, these counts are fitted to a double-power law function of the form, $N(>S)=N'/S' [(S/S')^\alpha+(S/S')^\beta]^{-1}$. This yields the best-fit parameters of $N'=220\pm40$~deg$^{-2}$, $S'=4.8\pm0.2$~mJy, $\alpha=10.1\pm1.7$, and $\beta=1.7\pm0.1$).
The counts in blank fields and the best-fit function are also shown in Fig.~\ref{fig:counts}.  The flux densities of the counts are scaled to 1.1\,mm flux density assuming a modified black body with spectral index ($\beta$) 1.5, dust temperature 35~K, and $z=2.5$, if needed.

The combined ADF22 counts show a factor of three to five excess at 0.4--2.5~mJy, compared to the best-fit function for blank fields (Fig.~\ref{fig:counts}). It is reported that at least 60\% of the ADF22A sources have $z_{\rm spec}\approx3.09$ and hence genuine proto-cluster members should account for the excess in ADF22A counts (\citealt{2015ApJ...815L...8U}; \citeyear{2017ApJ...835...98U}). Although the accurate redshift of ADF22 SMGs are mostly unknown, two of them have $z_{\rm spec}=3.09$ (H.~Umehata et al., in preparation, Table~\ref{table:catalog}).
Thus the excess in the combined ADF22 counts is expected to be similarly explained by the contribution from proto-cluster members. The results therefore indicate the over-abundance of dusty galaxies with high dust-enshrouded SFRs and elevated dust production in the proto-cluster on a 10 (projected) comoving Mpc scale. It is believed that abundant gas fueling via cosmic web filaments and/or frequent mergers in the center of the proto-cluster is the of cause high SFRs, and the accelerated assembly of stellar mass leads to the dust enrichment. 

One aim of our expansion of the 1~mm survey area was to investigate the transition of the level of dusty star-forming activity across the proto-cluster.
As we described above, the 3D structure traced by LAEs (\citealt{2005ApJ...634L.125M}) suggests that ADF22A corresponds to the exact intersection of the filamentary, large-scale structure (or ``cosmic web'' on 50 comoving-Mpc scale) while ADF22B encompasses the surrounding area (the ``filaments'' of the structure). As a result, the counts in ADF22A and ADF22B fields are in good agreement generally and hence the environmental dependence within the proto-cluster on this scale is not clear, while the overall ADF22B counts may lie below the ADF22A counts ($\approx$70--80\%) at $S_{\rm 1.1mm}\sim0.4-2.5$~mJy. While the brightest SSA22 SMG, with $S_{\rm 1.1mm}\ge$ 5~mJy, lies in the smaller area of ADF22A, this may be due to small-number statistics. 

Mapping a wider area of the proto-cluster and achieving a complete census of spectroscopic redshifts will be crucial next steps. It is also essential to understand the dust-obscured star formation in blank-field environments, which is still insufficient in many aspects. Wider and deeper contiguous mapping is required to derive number counts and luminosity functions correctly. Some recent and/or ongoing surveys (e.g.\  B.~Hatsukade et al, in preparation, and ALMACAL) are expected to push our understanding more in this area.

\subsection{ALMA SMGs and other rare populations}

As in the case of SMGs, X-ray AGNs and Ly$\alpha$ blobs (LABs), luminous, extended Ly$\alpha$ nebulae ($L_{\rm Ly\alpha}\gtrsim10^{43}$ ergs~s$^{-1}$; e.g.\  \citealt{2004AJ....128..569M}), are known to be rare populations seen in proto-cluster environments.
Over-abundances of both X-ray AGNs (\citealt{2009ApJ...691..687L}; \citeyear{2009MNRAS.400..299L}) and LABs (\citealt{2004AJ....128..569M}; \citeyear{2011MNRAS.410L..13M}) has been reported in the $z=3.09$ SSA22 proto-cluster and the connection to the large-scale structure has been discussed.
Additionally, the relation between SMGs and X-ray AGNs/LABs has been of interest for years, to investigate the co-evolution of galaxies and super-massive black holes (SMBHs) in the early Universe and/or to search for a heating source to generate the extended Ly$\alpha$ emission seen as LABs (e.g.\  \citealt{1998MNRAS.298..583I, 2009ApJ...700....1G}; \citealt{2013MNRAS.430.2768T}; \citealt{2016MNRAS.461.2944A}; \citealt{2016ApJ...832...37G}; \citealt{2016MNRAS.460.4075H}; \citealt{2017ApJ...834L..16U}; \citealt{2017ApJ...850..178A}).

Our contiguous ALMA map allows us to compare the positions of ALMA-identified SMGs with those of X-ray AGNs and LABs over a 20 arcmin$^{2}$ (71 comoving Mpc$^{2}$ at $z=3.09$) area.
There are five LABs listed in \citet{2004AJ....128..569M} in total within ADF22 in total, three of which (LAB11, LAB12, and LAB14) have SMG counterparts (see also \citealt{2015ApJ...815L...8U}) and two (LAB25 and LAB35) appear not to be associated with ADF22 SMGs.
The variety of outcomes suggests that LABs are not necessarily associated with a SMG, at least not at flux densities  bright enough to be detected in our map.

\citet{2016MNRAS.461.2944A} reported the elevated mean SFRs of X-ray AGNs located in the SSA22 proto-cluster core (up to a factor of $\approx$4.3) compared to the field, using pointed ALMA observations, and suggested that the growth of (AGN-host) galaxies is on average accelerated in the core.  Intruigingly, although six ADF22A SMGs at $z_{\rm spec}=3.09$ have X-ray AGNs (\citealt{2015ApJ...815L...8U}), none of ADF22B SMGs have a secure X-ray counterpart. Assuming that a significant fraction of ADF22B SMGs are at $z=3.09$, as discussed in \S\,4.1, this difference implies that the relative growth rate between galaxies and SMBHs depends on where the galaxies reside in large-scale cosmic structure.  


\subsection{Comparison between ALMA and AzTEC maps}

\begin{figure*}
  \begin{center}
\includegraphics[width=16.0cm]{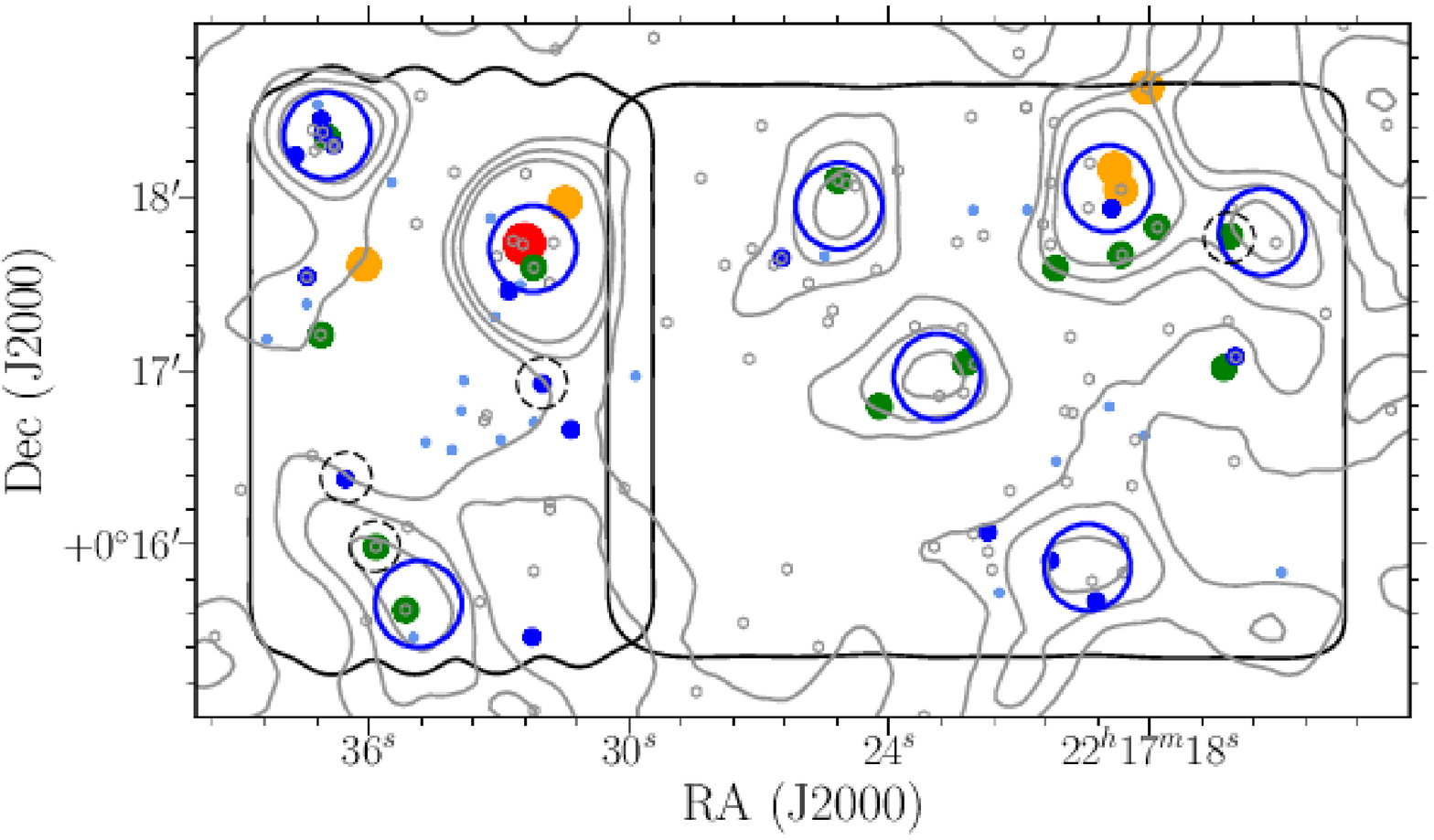}
  \end{center}
  \caption{%
The spatial distribution of the ALMA SMGs and AzTEC sources in the ADF22, including both ADF22A (\citealt{2017ApJ...835...98U}) and ADF22B.
Filled circles represent the ALMA SMGs, the size and color of which indicate their flux densities (red: $S_{\rm 1.1mm}>5$ mJy, orange: 2 mJy $<S_{\rm 1.1mm}\le5$ mJy, green: 1 mJy $<S_{\rm 1.1mm}\le2$ mJy, blue: $S_{\rm 1.1mm}\le 1.0$ mJy). Black contours delineate the ADF22 field area, as in Fig.~\ref{fig:lss}.
For reference, source candidates in supplementary catalogs are also displayed with light blue dots. Large blue open circles shows AzTEC sources (\citealt{2014MNRAS.440.3462U}), the diameters of which are equivalent to the beam FWHM of the AzTEC map ($d=30^{\prime\prime}$).
Dashed black circles shows the 870~$\mu$m ALMA pointing in ADF22 (\citealt{2016MNRAS.461.2944A}).
Gray contours represent the zero-level and 1.5, 3 $\sigma$ of the AzTEC 1.1\,mm map (here we adopt a fixed value, 1$\sigma$ = 0.7~mJy).
The ALMA SMGs appear to show a biased distribution which generally trace the sources and the diffuse, low-level emission in the AzTEC map. 
      }%
  \label{fig:dis_aztec}
\end{figure*}

\begin{figure}
  \begin{center}
\includegraphics[width=8.0cm]{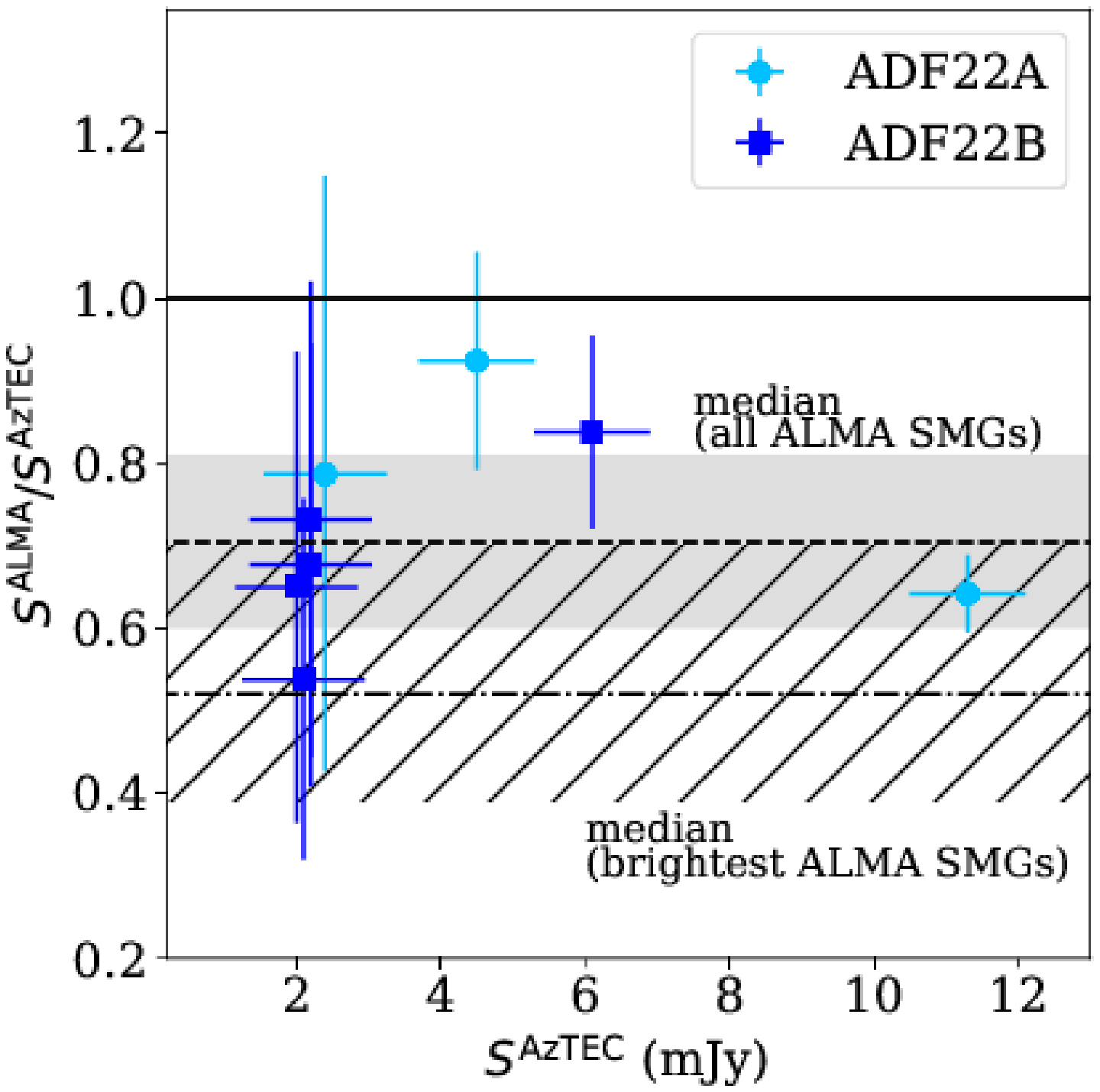}
  \end{center}
  \caption{%
Comparison of the flux measurements in the AzTEC and ALMA maps for eight AzTEC sources (\citealt{2014MNRAS.440.3462U}) located in the ADF22A or ADF22B.
$S^{\rm AzTEC}$ shows the flux density of the AzTEC sources and $S^{\rm ALMA}$ stands for the integration of that of the ALMA SMGs within the AzTEC beam. 
The dashed and dot-dashed lines with shaded/hatched regions show a median ratio of $S^{\rm ALMA}/S^{\rm AzTEC}$ and $S^{\rm ALMA}_{\rm bright}/S^{\rm AzTEC}$, respectively (here $S^{\rm ALMA}_{\rm bright}$ shows the flux of the brightest ALMA source within an AzTEC beam).
For some cases the sum of ALMA fluxes cannot account for the AzTEC flux, which suggests the additional contribution from extended and/or faint components. 
      }%
  \label{fig:flux_aztec}
\end{figure}

While an ALMA mosaic is the only way to achieve high angular resolution and high sensitivity, submm/mm surveys taken with single-dish telescopes can map much wider areas. Linking the two-types of maps is an important task. In this section, we compare the ALMA view from ADF22 with that of the AzTEC/ASTE map, where both surveys were deliberately matched in wavelength at 1.1\,mm\footnote{There is also a co-spatial SCUBA2 850 $\mu$m map (\citealt{2017MNRAS.465.1789G}). Since the difference in wavelength makes fair comparison difficult, we only focus on the AzTEC map here. }.  Here we investigate the relationship between the two maps, focusing on multiplicity, completeness, and flux density accuracy.

Fig.~\ref{fig:dis_aztec} shows the 1.1\,mm contours and source positions of the AzTEC map around the ADF22 field, compared to the spatial distributions and flux densities of ALMA SMGs. The combined ADF22 map supports findings based on the initial ADF22A map in several points (\citealt{2017ApJ...835...98U}). 
First, multiplicity is significant, as claimed by previous follow-up studies of the single-dish sources (e.g.\  \citealt{2007MNRAS.380..199I}; \citealt{2013ApJ...768...91H}; \citealt{2015ApJ...807..128S}). Two AzTEC sources in ADF22B have multiple ALMA SMGs within the 30$^{\prime\prime}$ beam, which means that five out of eight (63$_{-27}^{+37}$\%) AzTEC sources in ADF22 show multiplicity.  There is also a possible dependence on the flux density since the four brightest AzTEC sources ($S_{\rm 1.1mm, AzTEC}\ge2.2$~mJy) have multiple ALMA counterparts in all cases. 

Second, a number of moderately luminous ALMA SMGs ($S_{\rm 1.1mm, AzTEC}\sim1-2.5$~mJy) are located outside the AzTEC beam, though their flux densities are comparable to those of ALMA SMGs within the AzTEC beam. Six ALMA SMGs in ADF22B are found outside of the AzTEC source positions. Such a large fraction of the missed `mJy' population is also predicted by low completeness of the AzTEC source detection procedure (about 30\% at $S_{\rm 1.1mm, AzTEC}\sim2$~mJy; \citealt{2014MNRAS.440.3462U}). This result shows that an ALMA mosaic is crucial, even to obtain a complete picture of the `mJy' SMG population. 

As shown in Fig.~\ref{fig:dis_aztec}, the number of neighboring SMGs increases if we consider a slightly wider area than the AzTEC beam (the 30$^{\prime\prime}$ FWHM corresponds to $\approx$230~kpc at $z=3.09$).  We have shown that such {\it apparent} pairs/groups of SMGs on several hundred physical kpc scale is preferentially caused by {\it physically} associated pairs in ADF22A (\citealt{2017ApJ...835...98U}), which is suggestive of similar situation in the ADF22B. A complete census of spectroscopic redshifts will be necessary to confirm this relation across the ADF22 field.  We also note that diffuse emission in the AzTEC map appears to trace the distributions of ALMA SMGs below the 3.5$\sigma$ source detection threshold (\citealt{2014MNRAS.440.3462U}). In the case of ADF22B, all 17 ALMA SMGs reside in area where the AzTEC emission show positive values. The tendency suggests the utility of faint AzTEC emission in identifying individual SMGs and evaluating dust-obscured star-formation activity, to a certain degree at least.

For the eight AzTEC sources, we also check the reproducibility of the AzTEC flux density from summing the flux from counterpart ALMA SMGs.  Fig.~\ref{fig:flux_aztec} shows the fraction of the integrated ALMA flux compared to the deboosted AzTEC source flux (i.e., the flux densities corrected for boosting effect by random noise), considering only securely detected ALMA sources within the AzTEC beam. Although there are large uncertainties mainly from the AzTEC flux measurements, including an uncertainty of intrinsic AzTEC number counts in such a local overdense region, we cannot reproduce the AzTEC flux densities in some cases. The median, $S^{\rm ALMA}/S^{\rm AzTEC} =0.70_{-0.10}^{+0.11}$, is well below unity. We also measure a median ratio of $S^{\rm ALMA}_{\rm brightest}/S^{\rm AzTEC} =0.52_{-0.13}^{+0.18}$ (and $S^{\rm ALMA}_{\rm brightest}/S^{\rm ALMA} =0.89_{-0.23}^{+0.11}$), considering the brightest ALMA components. Thus while brightest ALMA SMGs dominate the summed contribution from all ALMA SMGs on average, the discrepancy between ALMA and AzTEC fluxes suggests that there is contribution from additional sources (e.g.\ extended and/or faint components). Since our untapered angular resolution is considerably higher than we chose (or would ideally choose), the ADF22B map is not sensitive to faint, extended emission.

\begin{figure}
  \begin{center}
\includegraphics[width=8.5cm]{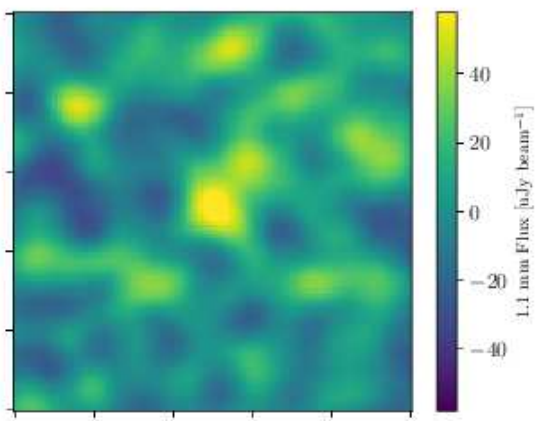}
  \end{center}
  \caption{%
The stacked ALMA 1.1\,mm image of 67 DRGs located in ADF22, which is $10^{\prime\prime}\times10^{\prime\prime}$ in size. The result shows that the DRGs have flux density $S_{\rm 1.1mm}=0.13\pm0.02$ mJy on average, and hence such color-selected galaxies in the vicinity of SMGs can account for some fraction of the AzTEC flux.
      }%
  \label{fig:stack}
\end{figure}

One possibility to explain the discrepancy is the contribution from {\it known} galaxy populations which would have relatively faint submm/mm emission. 
For instance, \citet{2014ApJ...780..115D} reported the detection of stacked 870~$\mu$m emission for color-selected galaxies, such as distant red galaxies (DRGs -- for which, e.g., \citealt{2005ApJ...632L...9K} and \citealt{2007ApJ...660L..77I} also report relevant stacking results), in the vicinity of ALMA-identified SMGs in ALESS, though the survey was designed for bright SMGs and so the sensitivity limit was relatively shallow (\citealt{2013ApJ...768...91H}). They also suggested that galaxies around bright SMGs may tend to be brighter at submm/mm (but see also \citealt{2016ApJ...833..195W}). 

In order to investigate whether such color-selected galaxies can account for the {\it missing} flux density for the AzTEC sources in ADF22, we perform a stacking analysis utilizing DRGs ($J-K>1.4$ and $K<24$; \citealt{2013ApJ...778..170K}).  As shown in Fig.~\ref{fig:dis_aztec}, a total of 81 DRGs are known in ADF22, fourteen of which are individually detected by ALMA.  We stacked the remaining 67 DRGs at their {\it Ks}-band positions on the ALMA maps, calculating the average flux density.  Since the two maps in ADF22A and ADF22B have different angular resolutions, we smoothed the ADF22B map to match the ADF22A map (1$^{\prime\prime}$ resolution).  The resultant stacked image is shown in Fig.~\ref{fig:stack}. The flux density measurement using {\sc casa}/{\sc imfit} shows that the ``1mm-faint'' DRGs have a flux density, $S_{\rm 1.1mm}=0.13\pm0.02$ mJy, on average.  Considering the fact that multiple DRGs without individual ALMA detection are generally encompassed within the AzTEC beam (two sources per AzTEC beam on average; Fig.~\ref{fig:dis_aztec}), the result suggests that such relatively massive, color-selected galaxies may account for some portion of the missed flux.  
For instance, two average DRGs, each positioned roughly an AzTEC beam FWHM from an AzTEC source, may account for $\approx 30$\% of the missed flux density on average or provide a elevated median ratio ($S^{\rm ALMA}/S^{\rm AzTEC} =0.82_{-0.16}^{+0.07}$).  
Thus our stacking analysis shows that color-selected galaxies can contribute to the flux densities of single-dish sources, even if they are not detected by ALMA individually at the current sensitivity limit.  We note that there are still other possibilities such as faint submm/mm galaxy populations without optical-to-NIR counterparts and extended dust emission.

\section{Summary}

The SSA22 proto-cluster at $z=3.09$ is one of the most remarkable and well-studied intergalactic structures in the early Universe. In order to uncover dust-obscured star-formation activity in the proto-cluster field, free from source confusion, we have obtained new ALMA data covering a contiguous 13~arcmin$^2$ region (ADF22B) at 1.1\,mm (band 6) towards the core of the known proto-cluster.  The 1.1\,mm continuum map thus obtained has a typical 1$\sigma$ sensitivity of 73 $\mu$Jy beam$^{-1}$ with a spatial resolution of 0$^{\prime\prime}$.5. Combined with our previous 1.1\,mm ALMA map (ADF22A; \citealt{2017ApJ...835...98U}), the two projects comprise a contiguous 20~arcmin$^2$ ALMA map in the SSA22 proto-cluster field (ADF22), one of the widest and deepest ALMA surveys to date.

We have obtained secure ($>5\sigma$) detections of 17 ALMA sources in ADF22B, making a total of 35 ALMA sources in ADF22, when combined with 18 ALMA sources found in ADF22A (\citealt{2017ApJ...835...98U}). The sources have flux densities, $S_{\rm 1.1mm}=0.43$--5.60 mJy; if we assume the SMGs lie at $z=3.09$, this corresponds to IR luminosities, $L_{\rm IR[8-1000\mu m]}=0.8_{-0.3}^{+0.2}\times 10^{12}$--$1.0_{-0.4}^{+0.3}\times 10^{13}L_\odot$ and to SFR$_{\rm IR}=80_{-40}^{+30}$ -- $1040_{-460}^{+340}$~$M_\odot$~yr$^{-1}$ (for a Chabrier IMF; considerably lower if we adopt the top-heavy IMF suitable for distant, dusty starbursts -- \citealt{2017MNRAS.470..401R}; \citealt{Zhang2018}). The bright, individual ALMA SMGs appear spatially resolved and have a median major axis $0.27_{-0.12}^{+0.05}$ arcsec (equivalent to 2.1$_{-0.9}^{+0.4}$~physical kpc at $z=3.09$), comparable to those of ALMA SMGs in blank fields. 

The cumulative number counts in ADF22 show a factor several excess compared to blank fields. This suggests that dust-obscured star formation and metal production are enhanced on a 10 comoving Mpc scale in the $z=3.09$ proto-cluster, though further follow-up ALMA spectroscopy is required to determine robust SMG redshifts and thus associations with the intergalactic structure.  

A comparison between SMGs and X-ray AGNs show apparent environmental variation within our survey. 
While SMGs are frequently associated with X-ray AGNs in ADF22A, no example of submm bright X-ray sources found in ADF22B, which implies that the relationships between the populations may differ as a function of density.

The comparison between our ALMA and AzTEC 1.1\,mm maps uncovers high multiplicity ratio of ALMA SMGs for an AzTEC source (63\%) and further local SMG over-densities around AzTEC source positions.  If a significant fraction of ALMA SMGs in ADF22B are genuine proto-cluster members, as is the case in ADF22A (\citealt{2017ApJ...835...98U}), then our results suggest that local SMG groups -- on scales of several hundred physical kpc -- account for the elevated dust-obscured star-formation activity seen across the proto-cluster.

The sum of the individually detected ALMA SMGs within the AzTEC beam is not sufficient to account for the AzTEC flux density, with a median flux density ratio, $S^{\rm ALMA}/S^{\rm AzTEC} = 0.70_{-0.10}^{+0.11}$. A stacking analysis of DRGs suggests that this particular galaxy population has an average flux density, $S_{\rm 1.1mm}=0.13\pm0.02$ mJy, and that such submm/mm-faint populations may contribute to the flux density of an AzTEC source, even if they are not detected individually by ALMA.

\begin{ack}
We thank the anonymous referee gratefully.
H.U.\ is supported by JSPS KAKENHI grant number 17K14252.
K.K.\ is supported by JSPS Grant-in-Aid for Scientific Research (S) JP17H06130. 
I.R.S.\ acknowledges support from the European
Research Council (ERC) Advanced Grant, DUSTYGAL (321334), a Royal Society/Wolfson Merit Award and STFC (ST/P000541/1).  R.J.I.\ acknowledges support from the ERC Advanced Grant, COSMICISM (321302).
This paper makes use of the following ALMA data:
ADS/JAO.ALMA\#2013.1.00162.S, ADS/JAO.ALMA\#2016.1.00580.S. ALMA is a partnership of ESO
(representing its member states), NSF (USA) and NINS (Japan), together
with NRC (Canada), NSC and ASIAA (Taiwan), and KASI (Republic
of Korea), in cooperation with the Republic of Chile. The Joint ALMA
Observatory is operated by ESO, AUI/ NRAO and NAOJ.
This research made use of APLpy, an open-source plotting package for Python (\citealt{2012ascl.soft08017R}).
\end{ack}

\appendix \section{Supplementary source catalog in ADF22}

Tentatively detected 1.1~mm sources in ADF22 are summarized in Table 3. 

\renewcommand{\arraystretch}{0.6} 
\begin{table*}
  \tbl{Supplementary Source Catalog}{%
  \begin{tabular}{ccccc}
  \hline
  (1) & (2) & (3) & (4) & (5)  \\
  ID & R.A.  & Dec.  & S/N & $S_{\rm 1.14mm}$ \\
      &           &          &        &  [mJy]               \\
      \hline
  & \multicolumn{3}{c}{ADF22B}  & \\
      \hline
ADF22B.18  &  22:17:14.95  &   +00:15:50:2  &   4.9  &   0.36 $\pm$ 0.04     \\ 
ADF22B.19  &  22:17:20.81  &   +00:17:55:6  &   4.8  &   0.67 $\pm$ 0.13    \\
ADF22B.20  &  22:17:22.07  &   +00:15:30:9  &   4.7  &   0.86 $\pm$ 0.16   \\
ADF22B.21  &  22:17:18.94  &   +00:16:47:5  &   4.7  &   0.58 $\pm$ 0.12   \\ 
ADF22B.22  &  22:17:18.12  &   +00:16:37:4  &   4.7  &   0.47 $\pm$ 0.10     \\ 
ADF22B.23  &  22:17:25.49  &   +00:17:39:5  &   4.6  &   0.39 $\pm$ 0.10   \\ 
ADF22B.24  &  22:17:21.46  &   +00:15:43:2  &   4.6  &   0.40 $\pm$ 0.10   \\ 
ADF22B.25  &  22:17:29.85  &   +00:16:58:0  &   4.5  &   0.74 $\pm$ 0.17    \\ 
ADF22B.26  &  22:17:20.15  &   +00:16:28:6  &   4.5  &   0.77 $\pm$ 0.16     \\
      \hline
  & \multicolumn{3}{c}{ADF22A}  & \\
      \hline
ADF22A.19 & 22:17:33.87 & +00:16:46.1  & 4.3 &  0.29 $\pm$ 0.06  \\
ADF22A.20 & 22:17:34.69 &  +00:16:35.2 & 4.3&  0.67 $\pm$ 0.09  \\
ADF22A.21 & 22:17:33.09 & +00:17:18.5  &  4.3 & 0.79 $\pm$ 0.08   \\
ADF22A.22 & 22:17:32.96 & +00:16:36.0  &  4.3 & 0.42 $\pm$ 0.07  \\
ADF22A.23 & 22:17:35.47 & +00:18:05.1  &  4.2 & 0.30 $\pm$ 0.06  \\
ADF22A.24 & 22:17:37.43   &+00:17:23.0  &   4.1 & 0.56 $\pm$ 0.08  \\
ADF22A.25 &  22:17:33.81 &+00:16:56.5  &   4.1 & 0.44 $\pm$ 0.08  \\
ADF22A.26 &  22:17:33.19& +00:17:52.7  &   4.1 & 0.63 $\pm$ 0.07 \\
ADF22A.27 &  22:17:37.18 &+00:18:32.0  &   4.0 & 0.25 $\pm$ 0.06  \\
ADF22A.28 &  22:17:32.50 &+00:17:29.5  & 4.0 & 0.48 $\pm$ 0.08  \\
ADF22A.29 &  22:17:38.35 &+00:17:10.8  &  4.0 & 1.12 $\pm$ 0.11 \\
ADF22A.30$^{\triangleright}$ &  22:17:32.19 &+00:16:42.0  &  4.7 & 0.29 $\pm$ 0.09 \\
ADF22A.31$^{\triangleright}$ &  22:17:34.97 &+00:15:27.6  &   4.6  & 0.41 $\pm$ 0.12  \\
ADF22A.32$^{\triangleright}$ &  22:17:34.08 &+00:16:32.6  &   4.5 & 0.35 $\pm$ 0.12  \\
      \hline
  \end{tabular}}\label{table:stable}
    \begin{tabnote}
(1) ID in this paper. 
(2) Right ascension in the wcs system.
(3) Declination in the wcs system.
(4) Signal-to-noise (S/N).
(5) Integral flux density measured with {\sc casa}/{\sc imfit}.
$^{\triangleright}$: ADF22A sources selected from a $0.7^{\prime\prime}$ map (`DEEP/HIRES' map; see \citealt{2017ApJ...835...98U}). The other ADF22A sources are selected from a  $1^{\prime\prime}$ map (`WIDE/LORES' map).
 \end{tabnote}
\end{table*}
\renewcommand{\arraystretch}{1}

\end{document}